\documentclass[letterpaper]{revtex4}

\usepackage{graphicx}

\begin{document}

\noindent
\large
\centerline{{\bf
Overlapping Resonances Interference-induced Transparency:}}
\centerline{{\bf
The $S_0 \to S_2/S_1$ Photoexcitation Spectrum of Pyrazine}}

\normalsize
\begin{quote}

{\bfseries Timur Grinev}

{\slshape \footnotesize Chemical Physics Theory Group, Department of Chemistry, and Center for Quantum Information and Quantum Control, University of Toronto, Toronto, Ontario M5S 3H6, Canada}

{\bfseries Moshe Shapiro}

{\slshape \footnotesize Department of Chemistry, University of British Columbia, Vancouver, British Columbia \\ V6T 1Z1, Canada, and Department of Chemical Physics, Weizmann Institute of Science, Rehovot 76100, Israel}

{\bfseries Paul Brumer}

{\slshape \footnotesize Chemical Physics Theory Group, Department of Chemistry, and Center for Quantum Information and Quantum Control, University of Toronto, Toronto, Ontario M5S 3H6, Canada}

%\newline $^*${\bf e-mail:} {\rm mshapiro@chem.ubc.ca}}~~
%$^\dagger${\bf e-mail:} {\rm pbrumer@chem.utoronto.ca}}

\vskip .2truein

The phenomenon of ``overlapping resonances interference-induced transparency'' (ORIT) is introduced and studied in detail for the $S_0 \to S_2/S_1$ photoexcitation of cold pyrazine (C$_4$H$_4$N$_2$). In ORIT a molecule becomes transparent at specific wavelengths due to interferences between envelopes of spectral lines displaying overlapping resonances. An example is the $S_2\leftrightarrow S_1$ internal conversion in pyrazine where destructive interference between overlapping resonances causes the $S_0 \to S_2/S_1$ light absorption to disappear at certain wavelengths. ORIT may be of practical importance in multi-component mixtures where it would allow for the selective excitation of some molecules in preference to others. Interference induced cross section enhancement is also shown.

\end{quote}
\setcounter{section}{0}

\section{Introduction}
In ``Electromagnetically Induced Transparency'' (EIT) \cite{Scully-Zubairy-1997,Shapiro-Brumer-2003,EIT-Harris-1,EIT-Harris-2,EIT-Harris-3,EIT-Harris-4,EIT-Ichimura,EIT-Takeoka}, one creates a photoabsorption transparency window at certain frequencies by applying a (``coupling'') laser operating at another set of frequencies. In this paper we investigate a related, though distinct, physical phenomenon, called ``overlapping resonances interference-induced transparency'' (ORIT), where the transparency occurs due to interference between material waves within a molecule. Though ORIT is known for small systems \cite{shapiro-1972,Shapiro-1998}, it has not been investigated for polyatomic molecules where overlapping resonances are far more ubiquitous. In the present paper we introduce the ORIT phenomenon for polyatomic molecules, and examine it in detail in the $S_0 \to S_2/S_1$ photoabsorption of pyrazine (C$_4$H$_4$N$_2$).

\section{The Pyrazine $S_0 \to S_2/S_1$ Photoexcitation}

The ultrafast dynamics of pyrazine is of longstanding interest. For example, in recent work where the computational tools used in this paper were developed, we studied pyrazine $S_2 \leftrightarrow S_1$ internal conversion, using a pre-excited superposition state in $S_2$ as a starting point \cite{Christopher-Pyrazine-1,Christopher-Pyrazine-2,Christopher-Pyrazine-3,Ioannis-Pyrazine-1}, or creating such superposition state in the $S_0 \to S_2/S_1$ ultrafast laser excitation \cite{Ioannis-Pyrazine-2,Our-Pyrazine-1}. In the course of these prior studies \cite{Christopher-Pyrazine-2} a detailed understanding of the vibronic structure and intramolecular dynamics of pyrazine was obtained.

We denote as intramolecular $| \kappa \rangle$  ``resonances'' the vibrational states of the noninteracting $S_2$ electronic state, and the projector onto this resonance manifold as $Q = \sum_\kappa | \kappa \rangle \langle \kappa |$. These resonances are coupled to the vibrational states of the $S_1$ electronic state, denoted as $| \beta \rangle$, with $P = \sum_\beta | \beta \rangle \langle \beta |$ being the projector onto the manifold of these states. When we ignore the coupling to the pyrazine triplet $T$ states \cite{PyrazineS1ToT1IntersystemCrossing}, the eigenstates of the excited $S_1 + S_2$ Hamiltonian, denoted as $| \gamma \rangle$, take into account the $S_1\leftrightarrow S_2$ couplings. We thus have that $P + Q = I = \sum_\gamma | \gamma \rangle \langle \gamma |$ where $I$ is the identity operator on the excited states manifold. Even in the neglect of rotations, the $S_0 \to S_2/S_1$ photoabsorption process becomes a formidable 24 modes vibrational problem.

We now examine conditions under which the (weak field) photoabsorption cross section $\sigma(E)$ given as \cite{Shapiro-1998,Shapiro-1993}
\begin{equation}
\sigma(E_\gamma) = \frac{4 \pi^2 \omega_{\gamma,g}}{c} \left| \langle \gamma | \mu | g \rangle \right|^2, \label{Cross-section-gamma-1}
\end{equation}
vanishes, where $| g \rangle$ is the ground vibrational state of $S_0$, $\mu$ is the transition-dipole operator, and $\omega_{\gamma,g} \equiv (E_\gamma - E_g)/\hbar$ is the excitation frequency. For $E_\gamma$ above the $S_2$ potential minimum, the full excited vibronic state $| \gamma \rangle$ is composed of both the $S_2$ and the $S_1$ vibrational states (remembering that the $S_1$ potential minimum lies below the $S_2$ potential minimum). It was previously established \cite{Christopher-Pyrazine-1,Christopher-Pyrazine-2,Christopher-Pyrazine-3,ChemPhysLett-2009} that $\langle \kappa | \mu | g \rangle$, the dipole matrix elements for the pure $S_0 \to S_2$ transitions, are an order of magnitude larger that $\langle \beta | \mu | g \rangle,$ the analogous matrix elements for the pure $S_0 \to S_1$ transitions. Hence, one can introduce the so-called ``doorway states" approximation, according to which,
\begin{equation}
\langle \gamma | \mu | g \rangle = \langle \gamma | (P + Q) \mu | g \rangle = \sum_\beta \langle \gamma | \beta \rangle \langle \beta | \mu | g \rangle + \sum_\kappa \langle \gamma | \kappa \rangle \langle \kappa | \mu | g \rangle \approx \sum_\kappa \langle \gamma | \kappa \rangle \langle \kappa | \mu | g \rangle. \label{MatrixElementExpansion-gamma}
\end{equation}
This yields, for $\sigma(E_\gamma)$:
\begin{equation}
\sigma(E_\gamma) = \frac{4 \pi^2 \omega_{\gamma,g}}{c} \left| \sum_\kappa \langle \gamma | \kappa \rangle \langle \kappa | \mu | g \rangle \right|^2. \label{Cross-section-gamma-2}
\end{equation}

The cross section in Eq. (\ref{Cross-section-gamma-2}) can be expanded as
\begin{equation}
\sigma(E_\gamma) = \frac{4 \pi^2 \omega_{\gamma,g}}{c} \left[ \sum_\kappa \left| \langle \gamma | \kappa \rangle \langle \kappa | \mu | g \rangle \right|^2 + \sum_{\kappa \ne \kappa'} \langle g | \mu | \kappa \rangle \langle \kappa' | \mu | g \rangle \langle \kappa | \gamma \rangle \langle \gamma | \kappa' \rangle \right]. \label{Cross-section-gamma-3}
\end{equation}
The first (positive) part of Eq. (\ref{Cross-section-gamma-3}),
\begin{equation}
\sigma^{\mathrm{diag}} (E_\gamma) = \sum_\kappa \frac{4 \pi^2 \omega_{\gamma,g}}{c} \left| \langle \gamma | \kappa \rangle \langle \kappa | \mu | g \rangle \right|^2 \equiv \sum_\kappa \sigma^{\mathrm{diag}}_\kappa (E_\gamma), \label{Cross-section-gamma-diag}
\end{equation}
is the diagonal contribution to $\sigma (E_\gamma)$, composed of individual contributions $\sigma^{\mathrm{diag}}_\kappa (E_\gamma)$ from the $| \kappa \rangle$ resonances, while
\begin{equation}
\sigma^{\mathrm{interf}} (E_\gamma) = \sum_{\kappa \ne \kappa'} \frac{4 \pi^2 \omega_{\gamma,g}}{c} \langle g | \mu | \kappa \rangle \langle \kappa' | \mu | g \rangle \langle \kappa | \gamma \rangle \langle \gamma | \kappa' \rangle \equiv \sum_{\kappa \ne \kappa'} \sigma^{\mathrm{interf}}_{\kappa,\kappa'} (E_\gamma) = \sum_{\kappa < \kappa'} 2 \, \sigma^{\mathrm{interf}}_{\kappa,\kappa'} (E_\gamma) \label{Cross-section-gamma-interf}
\end{equation}
is the (positive or negative) interference contribution to $\sigma (E_\gamma)$, composed of pairwise interference terms $\sigma^{\mathrm{interf}}_{\kappa,\kappa'} (E_\gamma)$, $\kappa \ne \kappa'$ ($\sigma^{\mathrm{interf}}_{\kappa,\kappa'} (E_\gamma)$ in Eq. (\ref{Cross-section-gamma-interf}) are assumed real). Each pairwise interference term $\sigma^{\mathrm{interf}}_{\kappa,\kappa'} (E_\gamma)$ is proportional to the product $\langle \kappa | \gamma \rangle \langle \gamma | \kappa' \rangle$, which is non-zero only if both $\langle \kappa | \gamma \rangle$ and $\langle \gamma | \kappa' \rangle$ are non-zero. When this happens, the $| \kappa \rangle$ and $| \kappa' \rangle$ resonances are said to {\it overlap}. The overlap is due to the common contribution of the full state $| \gamma \rangle$ (Ref. \cite{Christopher-Pyrazine-2,Our-Pyrazine-1}). In other words, for the pairwise interference contribution $\sigma^{\mathrm{interf}}_{\kappa,\kappa'} (E)$ between resonances $| \kappa \rangle$ and $| \kappa' \rangle$ to be non-zero at the particular energy $E = E_\gamma$, $| \kappa \rangle$ and $| \kappa' \rangle$ must overlap at this energy through one or more common states $| \gamma \rangle$.

The $\sigma (E_\gamma)$ cross section is closely related to the $S_2$ excited state population $P_{S_2} (t)$ \cite{Our-Pyrazine-1}, generated by CW pulse $\varepsilon(\omega) \approx \epsilon_a \delta(\omega - \omega_{\gamma,g})$ that excites only one state $| \gamma \rangle$. To obtain $P_{S_2} (t)$, consider the excited wave packet $| \Psi_a (t) \rangle$ on $S_1$ and $S_2$, produced by the one-photon transition from $S_0$ in the framework of first-order perturbation theory. For the CW case $| \Psi_a (t) \rangle$ takes the form:
\begin{equation}
| \Psi_a (t) \rangle = \frac{i \epsilon_a}{\hbar} \exp (-iE_\gamma t/\hbar) | \gamma \rangle \langle \gamma | \left[ \sum_\kappa \langle \kappa | \mu | g \rangle | \kappa \rangle \right]. \label{Psi_a-gamma-1}
\end{equation}
Taking into account Eq. (\ref{MatrixElementExpansion-gamma}), we see that the last sum of Eq. (\ref{Psi_a-gamma-1}) is $\mu | g \rangle \equiv | \Psi^e \rangle$ -- an excited wave packet on $S_2$. Thus,
\begin{equation}
| \Psi_a (t) \rangle = (i \epsilon_a/\hbar) \exp (-iE_\gamma t/\hbar) | \gamma \rangle \langle \gamma | \mu | g \rangle = (i \epsilon_a/\hbar) \exp (-iE_\gamma t/\hbar) | \gamma \rangle \langle \gamma | \Psi^e \rangle, \label{Psi_a-gamma-2}
\end{equation}
a result proportional to the action of the single-state propagator $\exp (-iE_\gamma t/\hbar) | \gamma \rangle \langle \gamma |$ on $| \Psi^e \rangle$. Noting that the population in the $S_2$ is given as
$P_{S_2} (t) = \langle \Psi_a (t) | Q | \Psi_a (t) \rangle$, we can use Eq. (\ref{Psi_a-gamma-2}) to obtain that
\begin{equation}
P_{S_2} = \frac{\epsilon_a^2}{\hbar^2} \langle \Psi^e | \gamma \rangle \langle \gamma | Q | \gamma \rangle \langle \gamma | \Psi^e \rangle = \frac{\epsilon_a^2}{\hbar^2} | \langle \gamma | \Psi^e \rangle |^2 \langle \gamma | Q | \gamma \rangle = \frac{\epsilon_a^2}{\hbar^2} | \langle \gamma | \Psi^e \rangle |^2 \left[ \sum_\kappa | \langle \kappa | \gamma \rangle |^2  \right]. \label{P_S_2_gamma}
\end{equation}
That is, the CW-generated $S_2$ population is time-independent, but $| \gamma \rangle$-dependent.

The $\sigma (E_\gamma)$ cross section of Eq. (\ref{Cross-section-gamma-1}) can also be written as the expectation value of the single-state projector $| \gamma \rangle \langle \gamma |$ in the $\mu | g \rangle \equiv | \Psi^e \rangle$ state,
\begin{equation}
\sigma(E_\gamma) = \frac{4 \pi^2 \omega_{\gamma,g}}{c} \langle g | \mu | \gamma \rangle \langle \gamma | \mu | g \rangle = \frac{4 \pi^2 \omega_{\gamma,g}}{c} \langle \Psi^e | \gamma \rangle \langle \gamma | \Psi^e \rangle = \frac{4 \pi^2 \omega_{\gamma,g}}{c} | \langle \gamma | \Psi^e \rangle |^2. \label{Cross-section-gamma-4}
\end{equation}
Comparing Eqs. (\ref{P_S_2_gamma}) and (\ref{Cross-section-gamma-4}) gives the following relationship between the CW-generated population at $E_{\gamma}$ and the cross section $\sigma (E_\gamma)$:
\begin{equation}
P_{S_2} = \frac{\epsilon_a^2 c}{4 \pi^2 \hbar^2 \omega_{\gamma,g}} \sigma (E_\gamma) \langle \gamma | Q | \gamma \rangle. \label{P_S_2_through_Sigma_E_gamma}
\end{equation}

\section{Computational Results for the Pyrazine $S_0 \to S_2/S_1$ Photoexcitation Cross Section}

\subsection{Coarse-Graining of the Pyrazine Vibronic Structure}

Pyrazine has 24 vibrational degrees of freedom and approximately 10$^{10}$ $| \gamma \rangle$ states \cite{Christopher-Pyrazine-2,Raab-1999} at $~\sim$2 eV above the $S_0 \to S_1$ vertical energy difference. For the computations to be numerically feasible, we replace the exact states with a set of approximate ``coarse-grained'' states \cite{Christopher-Pyrazine-2}. We thus divide the energy axis into 2000 bins $I_\alpha$ of size $\Delta_\alpha$, center energy $E_\alpha$ and density of states $\rho_\alpha$. The projector onto a coarse-grained state $| \alpha \rangle$ is defined as
$$| \alpha \rangle \langle \alpha | = (1/(\rho_\alpha \Delta_\alpha)) \sum_{\gamma \in I_\alpha} | \gamma \rangle \langle \gamma |,~~{\rm hence}~~ \sqrt{\rho_\alpha \Delta_\alpha} | \alpha \rangle \langle \alpha | \sqrt{\rho_\alpha \Delta_\alpha} =  \sum_{\gamma \in I_\alpha} | \gamma \rangle \langle \gamma | .$$
Thus, the coarse-grained state $| \alpha \rangle$ effectively replaces all the $| \gamma \rangle$ states in bin $I_\alpha$. Numerically, the weighted states $ |\overline{\alpha} \rangle  \equiv \sqrt{\rho_\alpha \Delta_\alpha} | \alpha \rangle $ and their overlaps with resonances $| \kappa \rangle$ are available through our iterative solution method for pyrazine (described in detail in Ref. \cite{Christopher-Pyrazine-2}).

The above coarse-grained description of the pyrazine vibronic structure is  fully adequate for vibronic femtosecond $S_2 \leftrightarrow S_1$ internal conversion dynamics \cite{Christopher-Pyrazine-2,Christopher-Pyrazine-3,Our-Pyrazine-1}. It involves a manifold of 76775 $ | \overline{\alpha} \rangle$ coarse-grained vibronic states, spanning the energy range of 2 eV above the $S_0 \to S_1$ vertical electronic transition. The density of $| \overline{\alpha} \rangle$ states is nonuniform with average vibronic energy separation of 2.6$\times$10$^{-5}$ eV = 0.21 cm$^{-1}$, which happens to coincide with some rotational energy spacings. Thus, if one considers vibronic $S_0 \to S_2$ excitation by a CW laser, which is narrow enough to resolve particular $| \overline{\alpha} \rangle$ states separately, one also needs to account for the rotational transitions and consider the full ro-vibronic spectrum. However, as discussed in the Appendix, if one assumes that pyrazine is initially cold, then consideration of vibronic levels only is sufficient.

To compute the cross section numerically, we thus replace $\langle \gamma | \kappa \rangle$ by $\langle \overline{\alpha} | \kappa \rangle$  and $E_\gamma$ by $E_\alpha$, throughout. When this is done, the cross section for the state $ | \overline{\alpha} \rangle$ is given [using Eq. (\ref{Cross-section-gamma-1})] as
\begin{equation}
\sigma(E_\alpha) =  \frac{4 \pi^2 \omega_{\alpha,g}}{c} \left| \langle \overline{\alpha} | \mu | g \rangle \right|^2 = \frac{4 \pi^2 \omega_{\alpha,g}}{c} \langle g | \mu | \overline{\alpha} \rangle \langle \overline{\alpha} | \mu | g \rangle \nonumber =  \frac{4 \pi^2 \omega_{\alpha,g}}{c} \langle \Psi^e | \overline{\alpha} \rangle \langle \overline{\alpha} | \Psi^e \rangle = \frac{4 \pi^2 \omega_{\alpha,g}}{c} | \langle \overline{\alpha} | \Psi^e \rangle |^2, \label{Cross-section-alpha-1}
\end{equation}
where $\omega_{\alpha,g} \equiv (E_\alpha - E_g)/\hbar$. In terms of $| \gamma \rangle$ states, this gives
\begin{equation}
\sigma(E_\alpha) = \sum_{\gamma \in I_\alpha} \frac{4 \pi^2 \omega_{\alpha,g}}{c} \langle \Psi^e | \gamma \rangle \langle \gamma | \Psi^e \rangle \approx \sum_{\gamma \in I_\alpha} \frac{4 \pi^2 \omega_{\gamma,g}}{c} \langle \Psi^e | \gamma \rangle \langle \gamma | \Psi^e \rangle = \sum_{\gamma \in I_\alpha} \sigma (E_\gamma). \label{Sigma_alpha_through_Sigma_gamma}
\end{equation}
Hence, $\sigma(E_\alpha)$ is approximately the cumulative sum of all the individual cross sections for all the $| \gamma \rangle$ in this bin $I_\alpha$. This is a reasonable approximation as long as the $I_\alpha$ bin size is small.

As in Eq. (\ref{Cross-section-gamma-2}), the coarse-grained cross section of Eq. (\ref{Cross-section-alpha-1}) can be decomposed into diagonal and interference contributions,
\begin{equation}
\sigma^{\mathrm{diag}} (E_\alpha) = \sum_\kappa \frac{4 \pi^2 \omega_{\alpha,g}}{c} \left| \langle \overline{\alpha} | \kappa \rangle \langle \kappa | \mu | g \rangle \right|^2 \equiv \sum_\kappa \sigma^{\mathrm{diag}}_\kappa (E_\alpha), \label{Cross-section-alpha-diag}
\end{equation}
and
\begin{eqnarray}
\sigma^{\mathrm{interf}} (E_\alpha) & = & \sum_{\kappa \ne \kappa'} \frac{4 \pi^2 \omega_{\alpha,g}}{c} \langle g | \mu | \kappa \rangle \langle \kappa' | \mu | g \rangle \langle \kappa | \overline{\alpha} \rangle \langle \overline{\alpha} | \kappa' \rangle \nonumber \\
& \equiv & \sum_{\kappa \ne \kappa'} \sigma^{\mathrm{interf}}_{\kappa,\kappa'} (E_\alpha) = \sum_{\kappa < \kappa'} 2 \, \sigma^{\mathrm{interf}}_{\kappa,\kappa'} (E_\alpha),\label{Cross-section-alpha-interf}
\end{eqnarray}
respectively ($\sigma^{\mathrm{interf}}_{\kappa,\kappa'} (E_\alpha)$ in Eq. (\ref{Cross-section-alpha-interf}) are real).

This result is analogous to that obtained earlier. That is, each pairwise interference term $\sigma^{\mathrm{interf}}_{\kappa,\kappa'} (E_\alpha)$ in the coarse-grained case is proportional to the product $\langle \kappa | \overline{\alpha} \rangle \langle \overline{\alpha} | \kappa' \rangle$, which is non-zero only if both $\langle \kappa | \overline{\alpha} \rangle$ and $\langle \overline{\alpha} | \kappa' \rangle$ are non-zero, \textit{i.e.}, the resonances $| \kappa \rangle$ and $| \kappa' \rangle$ overlap by means of the common coarse-grained state $| \overline{\alpha} \rangle$ \cite{Christopher-Pyrazine-2,Our-Pyrazine-1}. Only under such circumstances $\sigma^{\mathrm{interf}}_{\kappa,\kappa'} (E_\alpha)$ is non-zero.

\subsection{Computational Results for the $S_0 \to S_2/S_1$ Photoexcitation Cross Section}

In this study, transition dipole matrix elements $\langle \kappa | \mu | g \rangle$ are approximated by the corresponding Franck-Condon factors \cite{Christopher-Pyrazine-2}. One hundred and seventy six ``brightest'' $| \kappa \rangle$ resonances, having the largest Franck-Condon factors, were taken into account. These bright $| \kappa \rangle$ are numbered hereafter in order of ascending zero-order energy $E^0_\kappa$.

\begin{figure}[htp]
	\centering
	\includegraphics[width=0.8\textwidth, trim = 0 0 0 0]{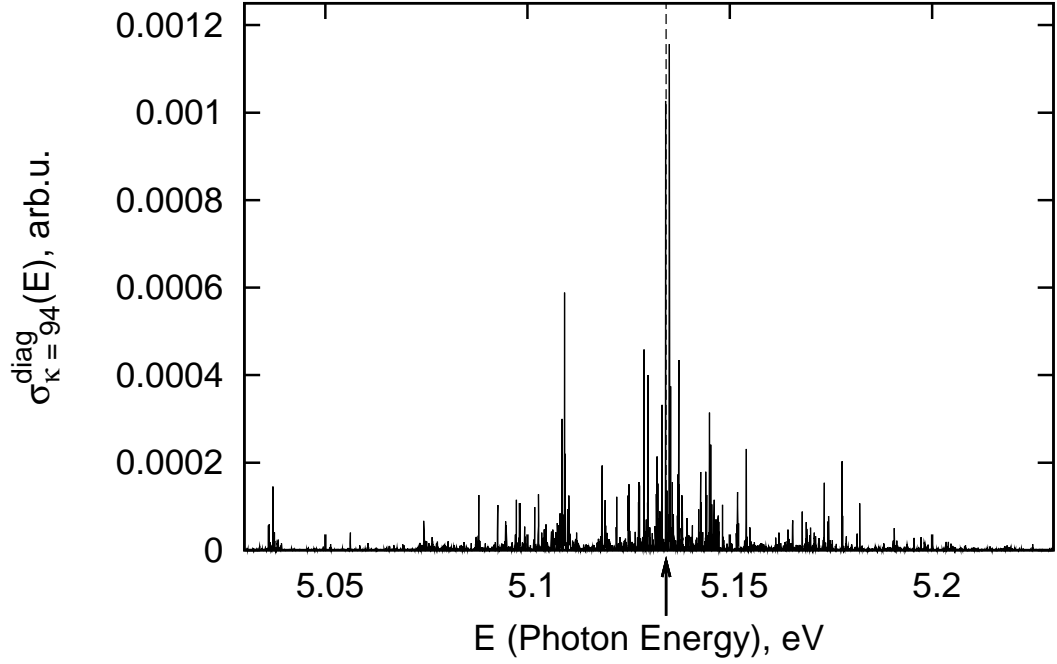}
	\renewcommand{\figurename}{FIG.}
	\caption{Cross section contribution $\sigma^{\mathrm{diag}}_{\kappa=94} (E)$, provided by the resonance $| \kappa = 94 \rangle$. Position of $E^0_{\kappa=94}$ is shown as a vertical dashed line and an arrow on the Energy axis.} \label{Figure-SingleResonance}
\end{figure}

The interference term in Eq. (\ref{Cross-section-alpha-interf}) can be positive or negative at any given $E_\alpha$, resulting in a constructive or destructive interference contribution to the photoabsorption cross section. The latter case is of particular interest, because it can lead to interference-induced transparency in the absorption profile. As an example, we consider the neighborhood of the particular resonance $| \kappa = 94 \rangle$. The diagonal term $\sigma^{\mathrm{diag}}_\kappa (E_\alpha)$ for this resonance is shown in Fig. \ref{Figure-SingleResonance}, together with the position of its zero-order energy, $E^0_{\kappa = 94} \approx$ 5.1342 eV. The energy of the resonance maximum, $E^{\max}_{\kappa=94}\approx$ 5.1350 eV, differs from $E^0_{\kappa=94}$ by a value $\Delta_{\kappa=94} \approx$ 8.0$\times$10$^{-4}$ eV (6.45 cm$^{-1}$), a resonance shift due to $S_1$--$S_2$ vibronic nonadiabatic coupling.

\begin{figure}[htp]
	\centering
	\includegraphics[width=0.7\textwidth, trim = 0 0 0 0]{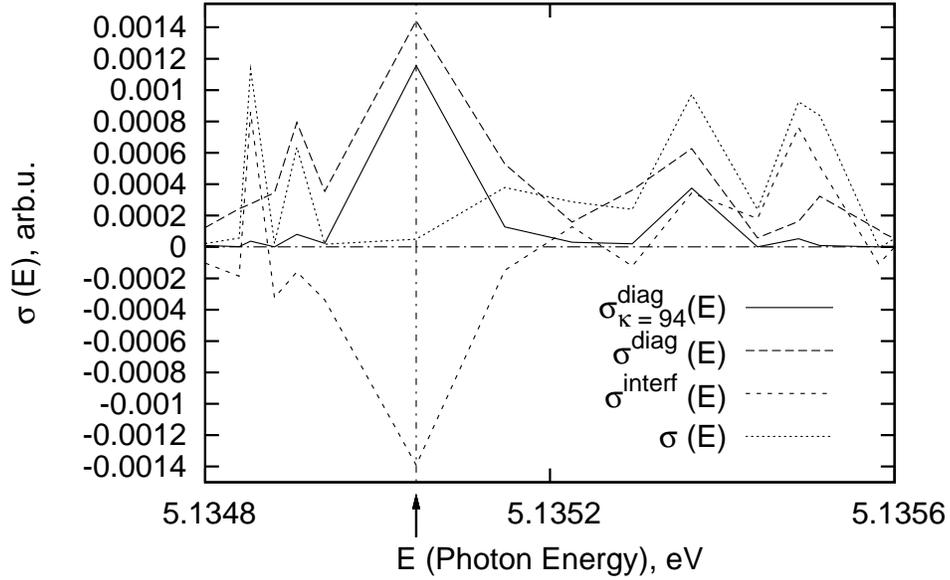}
	\renewcommand{\figurename}{FIG.}
	\caption{Cross section contribution $\sigma^{\mathrm{diag}}_{\kappa} (E)$ of resonance $| \kappa = 94 \rangle$, together with full diagonal cross section contribution $\sigma^{\mathrm{diag}} (E)$, full interference cross section contribution $\sigma^{\mathrm{interf}} (E)$, and full cross section $\sigma (E)$. Position of $E^{\max}_{\kappa=94}$ is shown as a vertical dash-dotted line and an arrow on the Energy axis.} \label{Figure-SngResPlusTotal}
\end{figure}

For the particular energy $E^{\max}_{\kappa=94}$, the contribution $\sigma^{\mathrm{diag}}_{\kappa=94} (E^{\max}_{\kappa=94})$ from the particular resonance $| \kappa = 94 \rangle$ dominates the full diagonal cross section $\sigma^{\mathrm{diag}} (E^{\max}_{\kappa=94})$, with $\sigma^{\mathrm{diag}}_{\kappa=94} (E^{\max}_{\kappa=94})/\sigma^{\mathrm{diag}} (E^{\max}_{\kappa=94}) \approx 0.80$ (see Fig. \ref{Figure-SngResPlusTotal}). One expects the total cross section $\sigma (E^{\max}_{\kappa=94})$ to be of the same order of magnitude, but in this particular case the total cross section $\sigma (E^{\max}_{\kappa=94})$ is actually much smaller than both $\sigma^{\mathrm{diag}} (E^{\max}_{\kappa=94})$ and $\sigma^{\mathrm{diag}}_{\kappa=94} (E^{\max}_{\kappa=94})$, leading to the local transparency. This transparency is due to the large negative (destructive) interference contribution $\sigma^{\mathrm{interf}} (E^{\max}_{\kappa=94})$. The result is shown in Fig. \ref{Figure-SngResPlusTotal}. Namely, the total cross section, composed of contributions from 176 bright $| \kappa \rangle$ resonances, is 23.4 times smaller at $E = E^{\max}_{\kappa=94}$ than the contribution from the one resonance $| \kappa = 94 \rangle$, an effect due to overlapping $| \kappa \rangle$ resonances.

\begin{figure}[htp]
	\centering
	\includegraphics[width=0.8\textwidth, trim = 0 0 0 0]{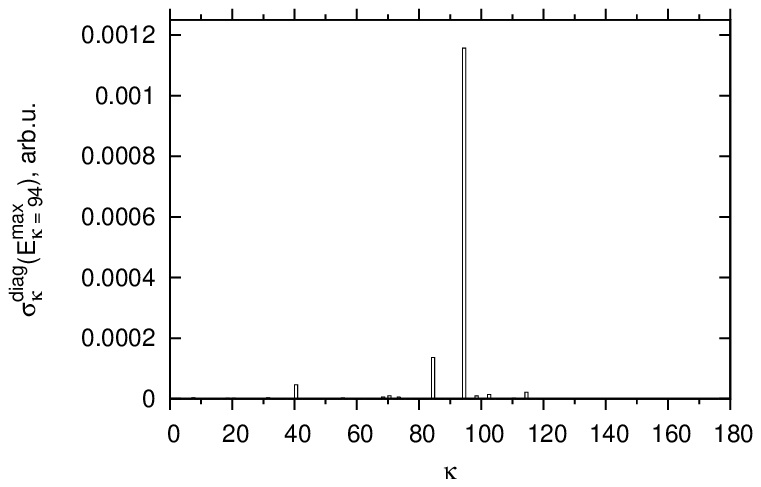}
	\renewcommand{\figurename}{FIG.}
	\caption{Individual cross section contributions $\sigma^{\mathrm{diag}}_{\kappa} (E)$ to $\sigma^{\mathrm{diag}} (E)$ at $E = E^{\max}_{\kappa=94}$, $\kappa$ = 1--176.}
\label{Figure-DiffResDiagContrib}
\end{figure}

It is instructive to analyze the structure of $\sigma^{\mathrm{diag}} (E^{\max}_{\kappa=94})$ and $\sigma^{\mathrm{interf}} (E^{\max}_{\kappa=94})$. The former is composed of 176 terms $\sigma^{\mathrm{diag}}_\kappa (E^{\max}_{\kappa=94})$, shown in Fig. \ref{Figure-DiffResDiagContrib}.
One can see that, out of overall 176, besides $\kappa$ = 94 itself, $\kappa$ = 40, 70, 84, 98, 102 and 114 contribute significantly to $\sigma^{\mathrm{diag}} (E^{\max}_{\kappa=94})$. This is a generic feature of pyrazine. That is, for any bright $|\kappa\rangle$ only several contributions are found to be important in $\sigma^{\mathrm{diag}} (E^{\max}_\kappa)$.
\begin{figure}[htp]
	\centering
	\includegraphics[width=0.8\textwidth, trim = 0 0 0 0]{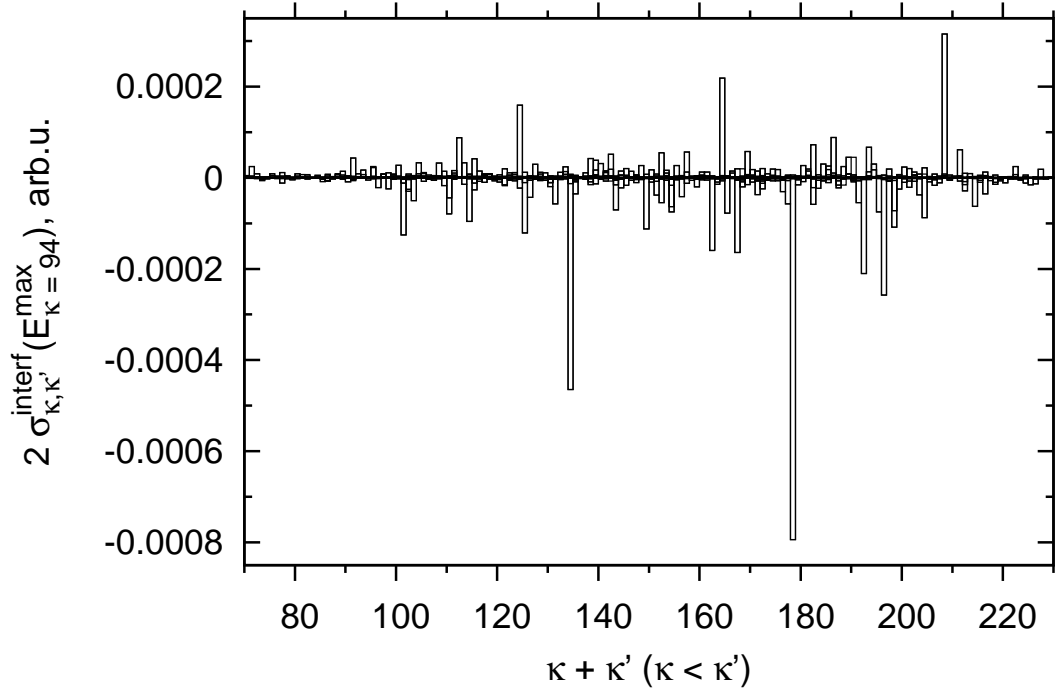}
	\renewcommand{\figurename}{FIG.}
	\caption{Pairwise interference cross section contributions $2\sigma^{\mathrm{interf}}_{\kappa,\kappa'} (E)$ to $\sigma^{\mathrm{interf}} (E)$ at $E = E^{\max}_{\kappa=94}$, $\kappa,\kappa'$ = 1--176 ($\kappa < \kappa'$).} \label{Figure-DiffResInterfContrib}
\end{figure}

For $\kappa$ = 1--176, there are 176(176--1)/2 = 15400 different pairwise interference contributions $\sigma^{\mathrm{interf}}_{\kappa,\kappa'} (E)$ in $\sigma^{\mathrm{interf}} (E)$. These interference contributions are presented in Fig. \ref{Figure-DiffResInterfContrib} for $E = E^{\max}_{\kappa=94}$, using a one-dimensional cumulative index $\kappa + \kappa'$ $(\kappa < \kappa')$ for simplicity. The most prominent contributions to the interference part of the cross section are seen to arise from pairs of resonances with largest (by absolute value) contributions to the diagonal part (Fig. \ref{Figure-DiffResDiagContrib}). Namely, the most significant pairwise interference contributions, seen in Fig. \ref{Figure-DiffResInterfContrib}, are the ones with $\kappa + \kappa'$ = 134 (\textit{i.e.} 40 + 94), 164 (70 + 94), 178 (84 + 94), 192 (94 + 98), 196 (94 + 102) and 208 (94 + 114). Thus, the overlap of $\kappa$ = 94 resonance with these several resonances is crucial for the destructive interference causing ORIT at $E = E^{\max}_{\kappa=94}$.

\begin{figure}[htp]
	\centering
	\includegraphics[width=0.7\textwidth, trim = 0 0 0 0]{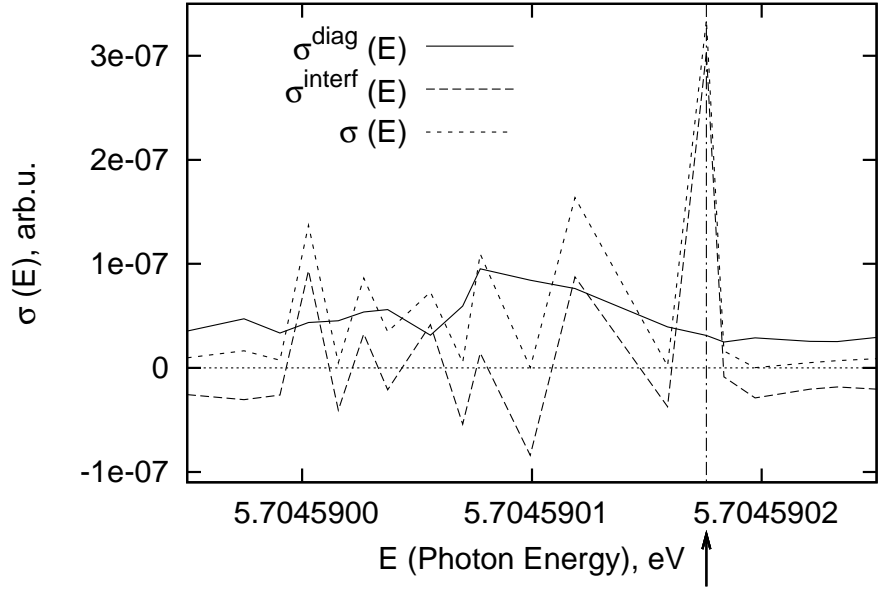}
	\renewcommand{\figurename}{FIG.}
	\caption{Full diagonal cross section contribution $\sigma^{\mathrm{diag}} (E)$, full interference cross section contribution $\sigma^{\mathrm{interf}} (E)$ and full cross section $\sigma (E)$. Energy of the coarse-grained state $| \overline{\alpha} = 66984 \rangle$, $E_{\alpha=66984} = 5.704590176$ eV, is shown as a vertical dash-dotted line and an arrow on the Energy axis.} \label{Figure-ConstructiveCase}
\end{figure}

In addition to ORIT, the interference term in Eq. (\ref{Cross-section-alpha-interf}) can be positive, thus leading to constructive interference, amplifying the total cross section $\sigma (E)$. One characteristic example of such interference is shown in Fig. \ref{Figure-ConstructiveCase} for the coarse-grained vibronic state $| \overline{\alpha} = 66984 \rangle$, having $E_\alpha = 5.704590176$ eV, for which total cross section $\sigma (E_\alpha)$ is 10.7 times larger than the diagonal cross section term $\sigma^{\mathrm{diag}} (E_\alpha)$; in other words, the interference term $\sigma^{\mathrm{interf}} (E_\alpha)$ here is 9.7 times larger than $\sigma^{\mathrm{diag}} (E_\alpha)$. While in case of ORIT $|\sigma^{\mathrm{interf}} (E_\alpha)| \le \sigma^{\mathrm{diag}} (E_\alpha)$ due to non-negativity of total cross section $\sigma (E_\alpha) = \sigma^{\mathrm{diag}} (E_\alpha) + \sigma^{\mathrm{interf}} (E_\alpha)$, there is no such a limitation on the absolute value of \textit{constructive} interference term with respect to diagonal term, as Fig. \ref{Figure-ConstructiveCase} shows.

\section{$S_0 \to S_2/S_1$ Photoexcitation Cross Section ORIT and CW-Generated $S_2$ Population}

The example of ORIT, presented and discussed above, is provided in terms of microscopic $\sigma (E)$ cross section. Physically, these effects should be observed in terms of full vibronic states $| \gamma \rangle$. However, the $S_2 \leftrightarrow S_1$ coupling, which is responsible for the very dense and rich $\sigma(E)$ structure, makes the experimental photoabsorption spectrum (obtained with finite spectral resolution and in the presence of environment) diffuse and continuous. This is the case for all pyrazine $S_0 \to S_2$ photoabsorption spectra, presented in literature \cite{Pyrazine-Experimental}, so that these available spectra do not allow comparison with our ORIT computation. So, instead, in this section we computationally confirm the relevance of ORIT in the cross section to the amount of CW-generated $S_2$ population, while the comparison with the experimental diffuse $S_0 \to S_2$ photoabsorption spectrum is presented in the next section below.

In terms of $| \overline{\alpha} \rangle$ states, the $S_2$ population is
\begin{equation}
P_{S_2} (t) = \frac{\epsilon_a^2}{\hbar^2} |\tau_\alpha(t)|^2 | \langle \overline{\alpha} | \Psi^e \rangle |^2 \langle \overline{\alpha} | Q  | \overline{\alpha} \rangle, \label{P_S_2_alpha-1}
\end{equation}
where $\tau_\alpha (t) = \exp (-i E_\alpha t/\hbar) \sin (\Delta_\alpha t/(2 \hbar))/(\Delta_\alpha t/(2 \hbar))$, so that $| \tau_\alpha (t) |^2 = [\sin (\Delta_\alpha t/(2 \hbar))/(\Delta_\alpha t/(2 \hbar))]^2$, which tends to be $\approx$ 1, if $|\Delta_\alpha t/(2 \hbar)| \ll 1$, or $|t| \ll 2\hbar/\Delta_\alpha$, which is the timescale of accuracy of coarse-grained description. So, for times $|t|$ small enough,
the CW-generated $S_2$ population is time-independent, but $| \overline{\alpha} \rangle$-dependent, and can be written as
\begin{equation}
P_{S_2} = \frac{\epsilon_a^2}{\hbar^2} | \langle \overline{\alpha} | \Psi^e \rangle |^2 \langle \overline{\alpha} |  Q  | \overline{\alpha} \rangle, \label{P_S_2_alpha-2}
\end{equation}
which is fully analogous to Eq. (\ref{P_S_2_gamma}) for the $| \gamma \rangle$ states.

\begin{figure}[htp]
	\centering
	\includegraphics[width=0.7\textwidth, trim = 0 0 0 0]{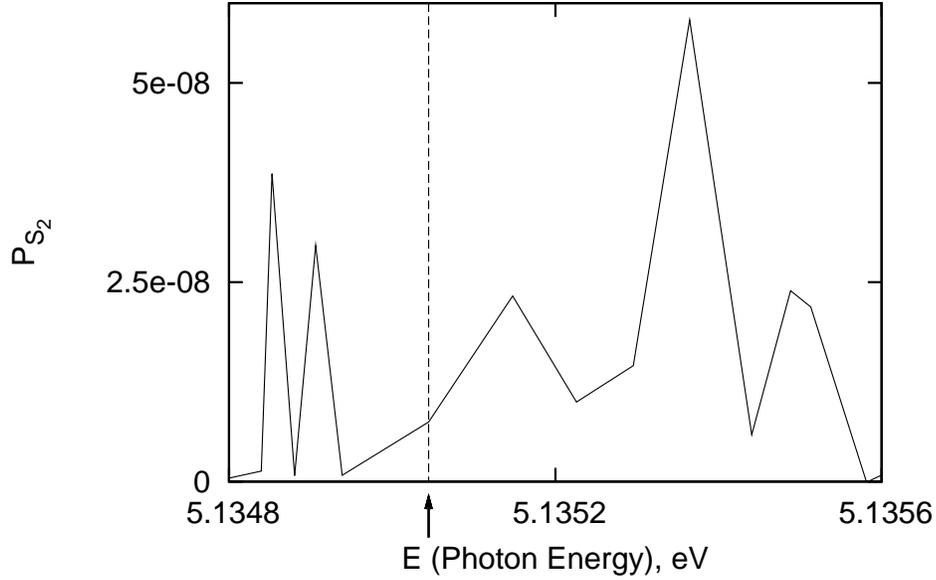}
	\renewcommand{\figurename}{FIG.}
	\caption{CW-generated $S_2$ population $P_{S_2}$, corresponding to the total cross section $\sigma (E)$ in Fig. \ref{Figure-SngResPlusTotal}. Position of $E^{\max}_{\kappa=94}$ is shown as a vertical dash-dotted line and an arrow on the Energy axis.} \label{Figure-P_S_2-CW}
\end{figure}

Comparison of Eqs. (\ref{P_S_2_alpha-2}) and (\ref{Cross-section-alpha-1}) provides the following relationship between the CW-generated $P_{S_2}$ and $\sigma(E_\alpha)$ using $| \overline{\alpha} \rangle$ states:
\begin{equation}
P_{S_2} = \frac{\epsilon_a^2 c}{4 \pi^2 \hbar^2 \omega_{\alpha,g}} \sigma (E_\alpha) \, \langle \overline{\alpha} |  Q  | \overline{\alpha} \rangle, \label{P_S_2_through_Sigma_E_alpha}
\end{equation}
which is similar to Eq. (\ref{P_S_2_through_Sigma_E_gamma}) in terms of $| \gamma \rangle$ states. The term $\langle \overline{\alpha} | Q | \overline{\alpha} \rangle = \sum_\kappa | \langle \kappa | \overline{\alpha} \rangle |^2$ in Eq. (\ref{P_S_2_through_Sigma_E_alpha}) is the incoherent sum, and is relatively slowly varying function of time. When $\sigma(E_\alpha)$ displays ORIT at specific energy $E = E_\alpha$ due to the destructive interference caused by the overlapping resonances, $P_{S_2}$ is also expected to be small. This expected $P_{S_2}$ behavior is fully confirmed computationally, using our pyrazine dynamics and control software from Ref. \cite{Our-Pyrazine-1}. This serves as a ``theoretical high resolution spectroscopy'', justifying the importance of the ORIT in the cross section for the computed $S_2$ population.
As an example, the computed $P_{S_2}$, corresponding to the total cross section $\sigma(E)$ in Fig. \ref{Figure-SngResPlusTotal} in the vicinity of $E = E^{\max}_{\kappa=94}$, is shown in Fig. \ref{Figure-P_S_2-CW}. The $P_{S_2}$ profile resembles the $\sigma(E)$ profile in Fig. \ref{Figure-SngResPlusTotal}; the distortion of $P_{S_2}$ shape from $\sigma (E)$ shape is mostly due to the varying $\langle \overline{\alpha} | Q | \overline{\alpha} \rangle$ factor in Eq. (\ref{P_S_2_through_Sigma_E_alpha}).

Beyond theoretical interest, ORIT can be of practical importance for certain applications. For example, as in ``EIT Spectroscopy''\cite{eilam}, one can selectively excite B in an A + B molecular mixture by a CW laser while leaving A in the ground state. To do so, one can use the local optical transparency of A at a certain laser frequency, when A does not absorb, provided that B absorbs well at this laser frequency.
\section{Comparison of the Computed $S_0 \to S_2$ Photoabsorption Spectrum with the Experimental Data}

Finally, we estimate the accuracy of our pyrazine vibronic structure and associated approximations by computing the $S_0 \to S_2$ photoabsorption spectrum $I(E)$ and comparing it with the available experimental data \cite{Pyrazine-Experimental}. To obtain $I(E)$, one can convolute the stick spectrum, composed of $| \langle \overline{\alpha} | \mu | g \rangle |^2$ values, with Lorentzian (as in Ref. \cite{Kuhl-Domcke}), having uniform FWHM$_E$ in energy domain:
\begin{equation}
I(E) \propto \omega_{E,g} \sum_\alpha | \langle \overline{\alpha} | \mu | g \rangle |^2 \frac{\Gamma/2}{(\omega_{E,g} - \omega_{\alpha,g})^2 + \Gamma^2/4}, \label{I_E_spectrum}
\end{equation}
where $\omega_{E,g} \equiv (E - E_g)/\hbar$, and $\Gamma = \mathrm{FWHM}_E/\hbar$. Alternatively, $I(E)$ can be computed using the Fourier transform of the autocorrelation function $C(t) = \langle \Psi(0) | \Psi(t) \rangle$, where $| \Psi(0) \rangle = \mu | g \rangle = | \Psi^e \rangle$, $| \Psi(t) \rangle = U(t) | \Psi(0) \rangle = \sum_\gamma \exp(-i E_\gamma t/\hbar) | \gamma \rangle \langle \gamma | \mu | g \rangle$, thus giving $C(t) \approx \sum_\alpha | \langle \overline{\alpha} | \mu | g \rangle |^2 \tau_\alpha (t) $ in terms of coarse-grained states $| \overline{\alpha} \rangle$. Then, as in Refs. \cite{Raab-1999,Pyrazine-Theory}, $I(E)$ is obtained as
\begin{equation}
I(E) \propto \omega_{E,g} \mathrm{Re} \int^\infty_0 dt \exp(-t/T_d) \, C(t) \exp(i \omega_{E,g} t), \label{I_E_Spectrum_FT_C_t}
\end{equation}
where $T_d$ is the damping parameter, giving the spectral broadening. 

\begin{figure}[htp]
	\centering
	\includegraphics[width=0.7\textwidth, trim = 0 0 0 0]{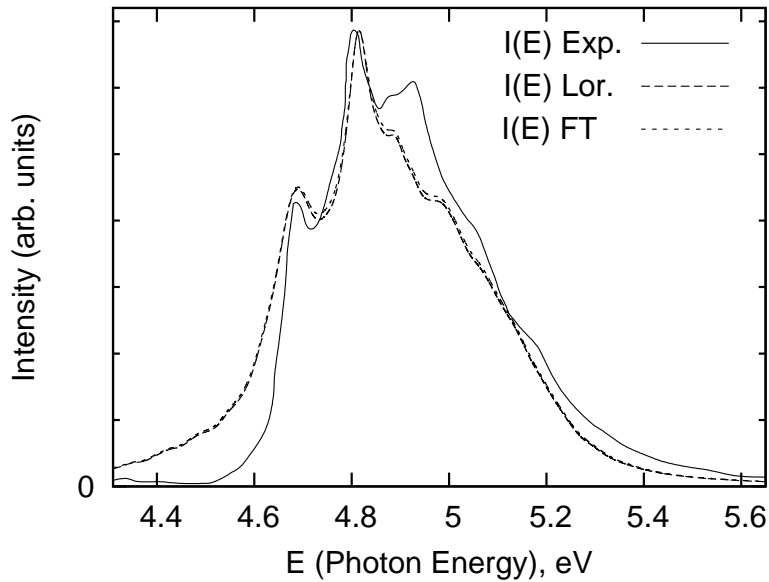}
	\renewcommand{\figurename}{FIG.}
	\caption{Experimental pyrazine $S_0 \to S_2$ photoabsorption spectrum (full line) from Ref. \cite{Pyrazine-Experimental} together with two calculated spectra (long dashed line and short dashed line). Here, FWHM$_E = $ 0.05 eV, $T_d = $ 25.0 fs.} \label{Figure-AbsorptionSpectrumCompVsExp}
\end{figure}

The spectrum $I(E)$, computed using Eq. (\ref{I_E_spectrum}) with FWHM$_E$ of 0.05 eV, and $I(E)$, computed using Eq. (\ref{I_E_Spectrum_FT_C_t}) with $T_d$ = 25.0 fs, are shown in Fig. \ref{Figure-AbsorptionSpectrumCompVsExp} along with the experimental spectrum of Yamazaki et al. \cite{Pyrazine-Experimental}. Fig. \ref{Figure-AbsorptionSpectrumCompVsExp} shows that the overall shape of the computed spectra (which are almost identical to each other) is in very good qualitative agreement with experiment (e.g., better than in Refs. \cite{Kuhl-Domcke, Schneider-Domcke}). Our computed spectra does lack the third peak near the maximum and somewhat overestimates absorption at low energies (before maximum), being less accurate than in Refs. \cite{Raab-1999,Pyrazine-Theory}. These quantitative differences likely reflect our use of the approximate coarse-grained $| \overline{\alpha} \rangle$ states, the simplifying restriction we impose \cite{Christopher-Pyrazine-2} to a total of 176 $S_2$ bright states, and the use of a phenomenological $\Gamma$ (or $T_d$) to describe the broadening of the $S_2$ states. These approximations, however, in no way affect the general physics of ORIT or of constructive spectral enhancement, which are the central focus of this paper.

\section{Summary and Conclusion}

In  conclusion, we have computationally demonstrated interference-induced transparency (and cross section enhancement) in a polyatomic molecule, using the $S_0 \to S_2/S_1$ photoabsorption cross section in pyrazine as an example. Specifically, transparency has been shown to arise from destructive quantum interferences associated with overlapping resonances \cite{shapiro-1972} which are optically accessible in the $S_0 \to S_2$ photoexcitation. In the specific case examined as an example, the destructive interference between several locally important resonances was found to be responsible for the local transparency effect. Interference induced cross section enhancement was also shown. These effects are expected to be ubiquitous in polyatomic molecules, a feature which has already been shown relevant to the coherent control of internal conversion \cite{Christopher-Pyrazine-3, Our-Pyrazine-1}.

\section{Acknowledgements}

The authors thank the Natural Sciences and Engineering Research Council of Canada and the NSF, under grant number CHE0848198, for funding.

\section*{Appendix: Consideration of Pyrazine Rotational Spectrum}

Pyrazine is a rigid aromatic system. For low rotational angular momentum quantum numbers, one can neglect rovibrational interaction and consider the molecular Hamiltonian as a sum of rotational and vibronic parts, being independent of one another. This gives the molecular wavefunction as a product of rotational and vibronic wavefunctions, and the molecular energy as a sum of rotational and vibronic energies governed by different quantum numbers.

Pyrazine is a nearly symmetric oblate top with rotational constants $\overline{B} = (A + B)/2 \approx$ 0.2 cm$^{-1}$, and $C \approx$ 0.1 cm$^{-1}$ \cite{Pyrazine-Rotational-1,Pyrazine-Rotational-2}.  Its rotational energy can be approximated as $E_R (J, K) = \overline{B} J (J + 1) + (C - \overline{B})K^2 \approx 0.2 J (J + 1) - 0.1 K^2$ cm$^{-1}$, where $J$ and $K$ are rotational angular momentum and its projection on the figure axis of the top, respectively ($J \ge K$). The symmetric top rotational selection rules \cite{Zare} allow only $|J'', K'' \rangle \to | J', K' \rangle$ transitions, belonging to $P$, $Q$, and $R$ branches, having $\Delta J \equiv J' - J'' = -1$, $0$, $+1$, respectively, with $\Delta K \equiv K' - K'' = 0$, $\pm 1$. In case of pyrazine, belonging to the point symmetry group $D_{2h}$, $S_0 (^1 \! A_g) \to S_2 (^1 \! B_{2u})$ transition dipole moment belongs to $B_{2u}$ symmetry (denoted $T_y$ or $y$ in the $D_{2h}$ character table). This transition dipole moment lies in molecular plane,  perpendicular to the figure axis of the top \cite{PyrazineSAndTStates,PyrazineConicalIntersection}, thus retaining only $\Delta K = \pm 1$ transitions in the $S_0 \to S_2$ rovibronic spectrum.

If the ground state is rotationally very cold, being comprised of only one lowest rotational state $| J'' = 0, K'' = 0 \rangle$, then only the $R$ branch exists. In the current case this gives only two $R$ branch transitions $| J'' = 0, K'' = 0 \rangle \to | J' = 1, K' = \pm 1 \rangle$. Both transitions have relative rotational line strengths of 1/2 \cite{Zare}, and the same $\Delta E_R \equiv E_R(J' = 1, K' = \pm 1) - E_R(J'' = 0, K'' = 0) \approx 0.3$ cm$^{-1}$, due to rotational $K$-doubling. Both transitions merge into the single line in the spectrum, uniformly shifting the vibronic spectrum by 0.3 cm$^{-1} = 3.72\times10^{-5}$ eV, which is very small in comparison with the energy of vibronic transition, starting here from 4.06 eV, $S_0 \to S_1$ vertical electronic excitation energy \cite{V-Batista-2006}. Considering the rotationally very cold ground state, one indeed can focus only on vibronic transitions.

\end{document}